# MULTI-VIEW IN LENSLESS COMPRESSIVE IMAGING

Hong Jiang, Gang Huang and Paul Wilford

Bell Labs, Alcatel-Lucent, Murray Hill, NJ 07974

*Abstract*—Multi-view images are acquired by a lensless compressive imaging architecture, which consists of an aperture assembly and multiple sensors. The aperture assembly consists of a two dimensional array of aperture elements whose transmittance can be individually controlled to implement a compressive sensing matrix. For each transmittance pattern of the aperture assembly, each of the sensors takes a measurement. The measurement vectors from the multiple sensors represent multi-view images of the same scene. We present theoretical framework for multi-view reconstruction and experimental results for enhancing quality of image using multi-view.

*Index Terms*— Compressed sensing, Image sensors, Image reconstruction

## I. INTRODUCTION

Lensless compressive imaging [1] is an effective architecture to acquire images using compressive sensing [2]. It consists of an aperture assembly and one or more sensors, but no lens is used. The transmittance of each aperture element is individually controllable. The sensors are used for taking compressive measurements. A compressive sensing matrix is implemented by adjusting the transmittance of the individual aperture elements according to the values of the sensing matrix. This architecture is distinctive in that the images acquired are not formed by any physical mechanism, such as a lens [3] or pinhole [4]. This results in the feature that no scene is out of focus, and the sharpness and resolution of images from the proposed architecture are only limited by the resolution of the aperture assembly, there is no blurring introduced by lens for scenes that are out of focus. Furthermore, the same architecture can be used for acquiring multimodal signals such as infrared, Terahertz [5] and millimeter wave images [6]. This architecture has application in surveillance [7].

Although only one sensor is considered in [1], the lensless compressive imaging architecture is well-suited for multi-view imaging because multiple sensors may be used in conjunction with one aperture assembly, see Figure 1. The cost of obtaining an additional viewpoint is simply that of adding a sensor to the device. For a given setting of transmittance, each sensor takes a measurement, and therefore, for a given sensing matrix, the sensors produce a set of measurement vectors simultaneously. Each measurement vector can be used to reconstruct an image independently without taking into consideration of other measurement vectors. However, although the images from multiple sensors are different, there is a high correlation between them, especially when the sensors are close to one another and when the scene is far away. The correlation between the images can be exploited to enhance the quality of the reconstructed images. Multi-view compressive imaging with lenses is considered in [8].

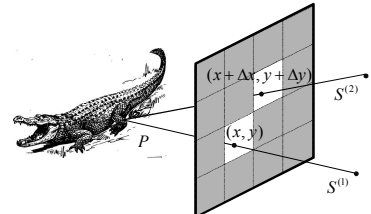

**Figure 1.** Lensless compressive imaging with two sensors.

Multiple sensors with one aperture assembly may be used in the following three ways:

**Multi-view**. In general, the measurement vectors from multiple sensors represent images of different views of a scene, creating multi-view images. This architecture allows a simple device to capture multi-view images simultaneously.

**Measurement increase**. When the scene is sufficiently far away, the measurement vectors from the sensors may be considered to be independent measurements of a common image and they may be concatenated into a larger set of measurements to reconstruct the common image. This effectively increases number of measurements that are taken for the image in a given duration of time.

**Higher Resolution**. When the scene is sufficiently far away, and when the sensors are properly positioned, the measurement vectors from the sensors may be considered to be the measurements made from a higher resolution image, and they may be used to reconstruct an image of the higher resolution than the number of aperture elements.

The purpose of this paper is twofold. First, we present a theoretical framework for reconstructing multi-view images by using joint reconstruction, which exploits the correlation between the multiple viewpoints. Second, we present experimental results to demonstrate how the multiple sensors can be used in each of the above three ways.

## II. MATHEMATICAL FORMULATION

### A. Virtual image and compressive measurements

As shown in Figure 1, for a given sensor, e.g., $S^{(1)}$, for each point $(x, y)$ on the aperture assembly, there is a ray starting from a point, $P$, on the scene, passing through the point $(x, y)$, and ending at the sensor. Denote by $r(x, y; t)$ the intensity of the unique ray associated with point $(x, y)$ on the aperture assembly at time $t$. An image $I(x, y)$ of the scene detected by the sensor is defined as the integration of the ray in a time interval $\Delta t$:

$$I(x,y) = \int_0^{\Delta t} r(x,y;t)dt. \quad (1)$$

The image in (1) is defined for each sensor, and it is called a virtual image because there is not an actual image formed by any physical mechanism. The virtual image can be pixelized by the aperture assembly. Let an aperture element be denoted by $E_{ij}$. Then the pixel value at the pixel $(i, j)$ is given by

$$I(i, j) = \iint_{E_{ij}} I(x, y)dxdy. \quad (2)$$

When the aperture assembly is programmed to implement a compressive sensing matrix, the transmittance of each aperture element is set to equal the value of the corresponding entry in the sensing matrix. Let the sensing matrix $A$ be a random matrix whose entries, $a_{mn}$, are random numbers between 0 and 1, and let $T^m(x, y)$ be the transmittance programmed according to row $m$ of $A$. Then the compressive measurements are given by

$$z_m = \iint T^m(x,y)I(x,y)dxdy = \sum_{i,j} a_{m,q(i,j)} I(i,j), \text{ or } z = A \cdot I, \quad (3)$$

where $q$ is mapping from a 2D array to a 1D vector, and $z$ is the measurement vector, $A$ is the sensing matrix and $I$ is the vector representation of the pixelized image $I(i, j)$.

It is well known [2] that image $I$ can be reconstructed from the measurements $z$ by, for example, solving the following minimization problem:

$$\min \|W \cdot I\|_1, \text{ subject to } A \cdot I = z, \quad (4)$$

where $W$ is some sparsifying operator such as total variation or framelets [7].

*B. Image decomposition*

We consider two sensors, $S^{(1)}$ and $S^{(2)}$, that are placed in a same plane parallel to the plane of aperture assembly, as in Figure 1. The sensors define two virtual images, $I^{(1)}(x, y)$ and $I^{(2)}(x, y)$. The geometry of the aperture assembly can be divided into two disjoint regions, $R_C^{(1)}$ and $R_D^{(1)}$, according to $S^{(1)}$. Region $R_C^{(1)}$ consists of the objects that can be seen by both $S^{(1)}$ and $S^{(2)}$; that is, the objects appearing in $R_C^{(1)}$ are common in both images $I^{(1)}(x, y)$ and $I^{(2)}(x, y)$. $R_D^{(1)}$ consists of the objects that can be only seen by $S^{(1)}$; that is, the objects appearing in $R_D^{(1)}$ can only be found in $I^{(1)}(x, y)$. $R_C^{(2)}$ and $R_D^{(2)}$ can be similarly defined as above by reversing the role of $S^{(1)}$ and $S^{(2)}$.

The definition of $R_C^{(1)}$ and $R_C^{(2)}$ also defines a one-to-one mapping between them. Referring to Figure 1, the points where the rays $\overrightarrow{PS^{(1)}}$ and $\overrightarrow{PS^{(2)}}$ intersects the aperture assembly are mapped into each other. The mapping is defined as

$$U^{12}: (x,y) \in R_C^{(1)} \mapsto (x+\Delta x, y+\Delta y) \in R_C^{(2)}$$
$$U^{21}: (x+\Delta x, y+\Delta y) \in R_C^{(2)} \mapsto (x,y) \in R_C^{(1)}, \quad (5)$$

where the relationship between $(x, y)$ and $(x+\Delta x, y+\Delta y)$ is shown in Figure 1.

Now the virtual images $I^{(k)}(x, y)$ can be decomposed according to $R_C^{(k)}$ and $R_D^{(k)}$ as follows

$$I^{(k)}(x,y) = I_C^{(k)}(x,y) + I_D^{(k)}(x,y), \quad k = 1, 2, \quad (6)$$

where $I_C^{(k)}(x, y)$ has support on $R_C^{(k)}$ and $I_D^{(k)}(x, y)$ has support on $R_D^{(k)}$. Furthermore, $I_C^{(1)}(x, y)$ and $I_C^{(2)}(x, y)$ are related through the following equations:

$$I_C^{(2)}(x,y) = I_C^{(1)}(U^{21}(x,y)), \ I_C^{(1)}(x,y) = I_C^{(2)}(U^{12}(x,y)). \quad (7)$$

The significance of Eq (6) is that the two virtual images are decomposed into three components: one component is common to both images, and the other two components are unique to each individual image, as shown below

$$I^{(1)}(x,y) = I_C(x,y) + I_D^{(1)}(x,y)$$
$$I^{(2)}(x,y) = I_C(U^{21}(x,y)) + I_D^{(2)}(x,y), \ I_C(x,y) = I_C^{(1)}(x,y). \quad (8)$$

Since $I_C(x, y)$ is common in both images, its reconstruction may make use of the measurements from both sensors, and therefore, its quality may be enhanced as compared to only one sensor is used.

Similarly, the pixelized images can be decomposed as

$$I^{(1)} = I_C + I_D^{(1)}, \quad I^{(2)} = U \cdot I_C + I_D^{(2)}. \quad (9)$$

In above, $U$ is a matrix that performs shift and interpolating functions to approximate the operation of mapping $U^{21}$ defined in (5). That is, $U \cdot I_C$ is a vector that approximates the pixelized $I_C(U^{21}(x, y))$, as given by

$$(U \cdot I_C)(q(i,j)) \approx \iint_{E_{ij}} I_C(U^{21}(x,y))dxdy. \quad (10)$$

*C. Joint reconstruction*

The vector components $I_C$, $I_D^{(1)}$ and $I_D^{(2)}$ may be jointly reconstructed from the two measurement vectors, $z^{(1)}$ and $z^{(2)}$, made from the two sensors, $S^{(1)}$ and $S^{(2)}$. Let $A$ be the sensing matrix with which the measurements $z^{(1)}$ and $z^{(2)}$ are made. Then the optimization problem to solve is

$$\min \|W \cdot I_C\|_1 + \frac{\sigma}{2} \sum_{k=1}^{2} \|W \cdot I_D^{(k)}\|_1, \text{ subject to}$$
$$A \cdot I_C + A \cdot I_D^{(1)} = z^{(1)}, \ A \cdot U \cdot I_C + A \cdot I_D^{(2)} = z^{(2)}. \quad (11)$$

In (11), $\sigma > 0$ is a normalization constant to account for the areas of the four regions $R_C^{(k)}$ and $R_D^{(k)}$, $k = 1, 2$. The significance of the joint reconstruction (11) lies in the fact that there are only three unknown components in (11) with two constraints (given by $z^{(1)}$ and $z^{(2)}$), as opposed to four unknown components with same number of constraints if the images are reconstructed independently from (4). Typically, $I_C$ has much more nonzero entries than that of $I_D^{(1)}$ and $I_D^{(2)}$, hence the number of unknowns is reduced by almost a half.

*D. Measurement increase*

When the scene is sufficiently far away, the images from the two sensors are approximately the same, except for a shift equal to the distance $d$ between two sensors. Therefore, the common region $R_C^{(k)}$ covers the entire aperture assembly except for a border of width $d$. Consequently, compared to the common image $I_C$, the images $I_D^{(1)}$ and $I_D^{(2)}$ have small energy. This implies that problem (11) is mainly a problem for the common image $I_C$, while using two measurement vectors $z^{(1)}$ and $z^{(2)}$, twice as many measurements as when each of the images, $I^{(1)}$ and $I^{(2)}$, is reconstructed independently in (4). For this reason, multiple sensors may be considered as taking independent measurements for a same image if the scene is sufficiently far away. This can be used as a mechanism to increase the number of measurements taken during a given time duration.

*E. Higher resolution*

For sufficiently far away scenes, multiple sensors may also be used as a mechanism to improve the resolution of the common image $I_C$. If the distance, $d$, between two sensors is a non-integer multiple of the size of the aperture elements, then $I^{(1)}$ and $I^{(2)}$ can be considered as two down-sampled images of a higher resolution image. The joint reconstruction can therefore be used to create a higher resolution image.

Specifically, equation (8) can be rewritten as

$$I^{(1)}(x,y) = I_C(x,y) + I_D^{(1)}(x,y),$$
$$I^{(2)}(x,y) = I_C(x-\Delta x, y-\Delta y) + I_D^{(2)}(x,y). \quad (12)$$

If the distance $d$ between two sensors is a non-integer multiple of the size of the aperture elements, then there is no overlapping of grid points $(x-\Delta x, y-\Delta y)$ with the grid points $(x,y)$. Therefore, equation (12) shows that images $I^{(1)}$ and $I^{(2)}$ comprise different sampling of the same image $I_C$, i.e., $I^{(1)}$ samples $I_C$ at points $(x,y)$, while $I^{(2)}$ samples $I_C$ at points $(x-\Delta x, y-\Delta y)$. Consequently, the measurement vectors $z^{(1)}$ and $z^{(2)}$ can be used to reconstruct the image $I_C$ at both grid points $(x,y)$ and $(x-\Delta x, y-\Delta y)$. This results in an image $I_C$ that has a higher resolution than given by the aperture elements.

## III. EXPERIMENT

A lensless compressive imaging prototype with two sensors [1] is shown in Figure 2. It consists of a transparent monochrome liquid crystal display (LCD) screen and two photovoltaic sensors enclosed in a light tight box. The LCD screen functions as the aperture assembly while the photovoltaic sensors measure the light intensity. The photovoltaic sensors are tricolor sensors, which output the intensity of red, green and blue lights.

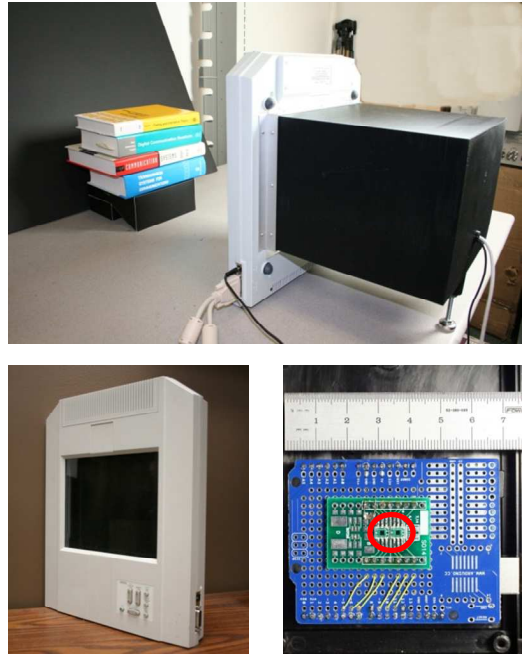

**Figure 2.** Prototype device. Top: lab setup. Bottom left: the LCD screen as the aperture assembly. Bottom right: the sensor board with two sensors, indicated by the red circle.

The LCD panel is configured to display 302x217 = 65534 black or white squares. Each square represents an aperture element with transmittance of a 0 (black) or 1 (white). A Hadamard matrix of order $N$=65536 is used as sensing matrix, which allows a total number of 65534, corresponding to the total number of pixels in the image, independent measurements to be made by each sensor. In our experiments, we only make a fractional of the total number of measurements. We express the number of measurements taken and used in reconstruction as a percentage of the total number of pixels. For example, 25% of measurements means 16384 measurements are taken and used in reconstruction, which is a quarter of the total number of pixels, 65534. In each experiment, a set of measurements is obtained by each sensor simultaneously. The two sensors are placed such that there is almost no vertical offset, and there is a horizontal offset of approximately 3.5 pixels.

*A. Measurement increase*

We compare the quality of images by individual and joint reconstructions in Figure 3, which is composed of six images, arranged in two columns and three rows. On the top row, the two images are reconstructed by Eq (4) using 12.5% (left) and 25% (right) of measurements taken from sensor 1 only. In the middle row, two images are the same; it is reconstructed by Eq (11) using 12.5% of measurements from each of the two sensors (for a combined 25%). On the bottom row, the two images are reconstructed by Eq (4) using 12.5% (left) and 25% (right) of measurements taken from sensor 2 only. We can make a couple of observations from Figure 3. First, as expected, the images using 25% measurements from one sensor only are clearly better than the images using 12.5% measurements from one sensor only. That is, an image on the right column, top or bottom row, is better than an image on the left column, top or bottom row. Second, the image from joint reconstruction using measurements from both sensors is better than images using 12.5% measurements from one sensor only, and as good as the images using 25% measurements from one sensor only, i.e., in the left column, the middle image is better than top and bottom; in the right column, all three images are similar. In reconstructing the image in the middle row, although a total of 25% of measurements are used, these measurements are taken in a time interval during which each sensor only takes 12.5% of measurements.

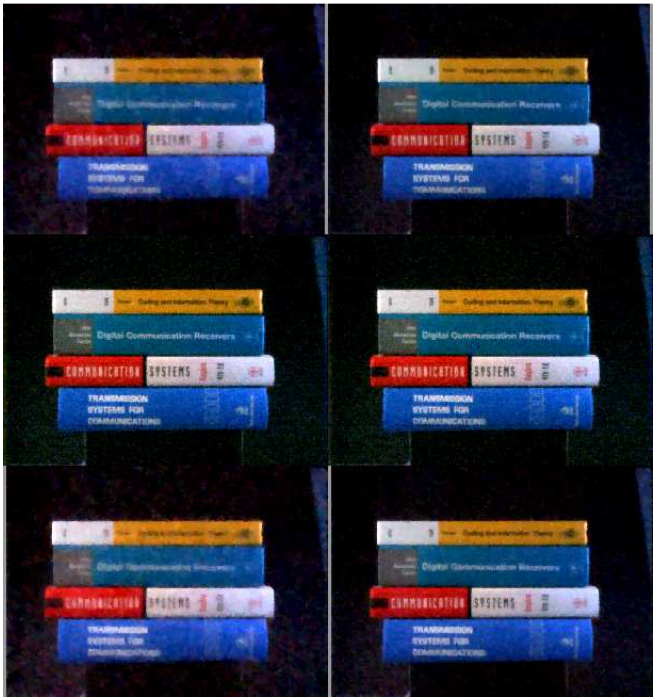

**Figure 3**. Reconstruction using measurements from two sensors.

*B. Higher resolution*

In Figure 4, the top and bottom images are reconstructed individually by Eq (4) using 25% of measurements taken from each of sensor 1 and sensor 2, respectively. The middle image is reconstructed using joint reconstruction to a higher resolution, 604x217, by using 25% measurements from each of two sensors, taking the advantage that there is a 3.5 pixels horizontal offset between the two sensors. It is evident that the image in the middle is sharper due to twice the horizontal resolution.

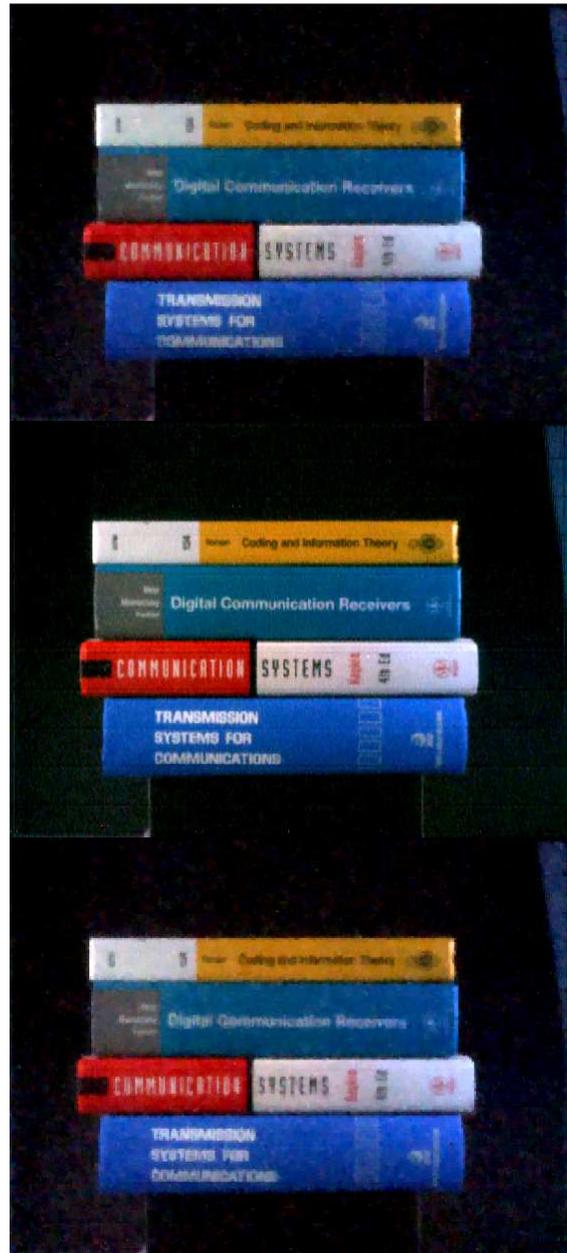

**Figure 4**. Reconstruction to higher resolution using measurements from two sensors.